\documentclass[aps,prl,twocolumn,groupaddress]{revtex4}
\usepackage{amsmath}
\usepackage{amsfonts}
\usepackage{amssymb}
\usepackage{epsfig}
\usepackage{graphicx}
\newcommand{\be}{\begin{equation}}
\newcommand{\ee}{\end{equation}}

\begin{document}

\title{A novel approach to numerical measurements of the configurational 
entropy in supercooled liquids}

\author{Ludovic Berthier}
\affiliation{Laboratoire Charles Coulomb, UMR 5221, CNRS and Universit\'e
Montpellier 2, Montpellier, France}

\author{Daniele Coslovich}
\affiliation{Laboratoire Charles Coulomb, UMR 5221, CNRS and Universit\'e
Montpellier 2, Montpellier, France}

\date{\today}

\begin{abstract}
The configurational entropy is among the key observables to
characterize experimentally the formation of a glass. Physically, 
it quantifies the multiplicity of metastable states in which an amorphous
material can be found at a given temperature, and its temperature 
dependence provides a major thermodynamic signature of the glass 
transition, which is experimentally accessible. Measurements of
the configurational entropy require, however, some approximations
which have often led to ambiguities and contradictory results. Here we
implement a novel numerical scheme to measure the configurational
entropy $\Sigma(T)$ in supercooled liquids, using a direct
determination of the free energy cost to localize the system within a
single metastable state at temperature $T$. For two prototypical
glass-forming liquids, we find that $\Sigma(T)$ disappears
discontinuously above a temperature $T_c$, which is slightly lower than
the usual estimate of the onset temperature for glassy dynamics. This
observation is in good agreement with theoretical expectations, but
contrasts sharply with alternative numerical methods. While the
temperature dependence of $\Sigma(T)$ correlates with the glass
fragility, we show that the validity of the Adam-Gibbs relation
(relating configurational entropy to structural relaxation time)
established in earlier numerical studies is smaller than previously thought,
potentially resolving an 
important conflict between experiments and simulations.
\end{abstract}


%05.10.-a 	Computational methods in statistical physics and nonlinear dynamics (see also 02.70.-c in mathematical methods in physics)

%05.20.Jj 	Statistical mechanics of classical fluids (see also 47.10.-g General theory in fluid dynamics)

%64.70.Q- 	Theory and modeling of the glass transition 

\maketitle

\section{Introduction}

The configurational entropy (or complexity) plays an important role 
in descriptions of the glass transition because it quantifies 
the temperature evolution of the free energy landscape accompanying  
changes in thermodynamic and dynamic properties of supercooled 
liquids. It represents both a major experimental signature of the 
glass transition~\cite{ediger_supercooled_1996} and a 
fundamental quantity within 
a number of theoretical approaches~\cite{berthier_theoretical_2011}.

The configurational entropy $\Sigma(T)$ is traditionally measured 
by subtracting a `vibrational' contribution to the total entropy
of the system: $\Sigma(T) \simeq S_{\rm tot}(T) - S_{\rm vib}(T)$. 
While $S_{\rm tot}(T)$ is well-defined, the vibrational 
contribution requires some approximation. 
Experiments~\cite{ediger_supercooled_1996,kauzmann_nature_1948,richert_dynamics_1998,martinez_thermodynamic_2001,angell_specific_2002}
use for instance the entropy of the crystalline or glass 
states to estimate $S_{\rm vib}(T)$. In simulations, 
the above decomposition relies on the assumption that the system 
vibrates around a given `state', further assumed to 
be equivalent to a local energy minimum, or inherent 
structure~\cite{stillinger_hidden_1982}. 
A thermodynamic formalism was developed to
determine numerically the configurational entropy, and applied to 
a large number of 
models~\cite{sciortino_inherent_1999,Scala_nature_2000,Sastry_2001,mossa_dynamics_2002,angelani_configurational_2007}.  
These studies have additionally revealed that the 
Adam-Gibbs relation~\cite{adam_temperature_1965} 
\begin{equation}
\tau_\alpha(T) \approx \tau_0 \exp \left( \frac{A}{T \Sigma(T)} \right),
\label{agr}
\end{equation}
between $\Sigma(T)$ and the structural relaxation 
time $\tau_\alpha(T)$ is obeyed over a broad temperature window.
In Eq.~(\ref{agr}), $A$ is an energy scale
and $\tau_0$ a microscopic timescale. Equation 
(\ref{agr}) is an important relation for supercooled liquids, as
its validity would directly establish that the viscosity increase
near the glass transition is caused by the temperature evolution 
of a complex free energy landscape. 
  
Available numerical methods are however not fully satisfactory
from both theoretical and experimental viewpoints. 
Firstly, the identification of metastable states 
with energy minima within the inherent structure formalism 
has been questioned~\cite{berthier_theoretical_2011,biroli_inherent_2000}. 
Because energy minima exist at all $T$, the inherent
structure $\Sigma(T)$ exists at arbitrarily high temperatures, 
where the free energy landscape is in fact featureless. 
In theoretical approaches~\cite{berthier_theoretical_2011,Wolynes_1997}, 
$\Sigma(T)$ is the entropic contribution stemming from
the multiplicity of metastable states proliferating at low $T$.
While this definition is also plagued by 
ambiguities, as discussed below, specific calculations show that 
$\Sigma(T)$ appears discontinuously below a temperature corresponding 
(within mean-field approximations) to the mode-coupling transition 
temperature~\cite{berthier_theoretical_2011,cavagna_supercooled_2009}. 
Secondly, the Adam-Gibbs relation in Eq.~(\ref{agr})
was numerically found to be valid over the 
entire supercooled regime~\cite{Scala_nature_2000,Sastry_2001,mossa_dynamics_2002,sengupta_dependence_2011}.
Experiments report instead that it only holds at low temperatures 
below the mode-coupling temperature~\cite{richert_dynamics_1998}, in a regime not accessible 
in simulations. These experimental findings are physically 
sensible because it is only at such low temperatures that the free energy 
landscape can possibly control the dynamics, but they directly 
contradict simulations.

\section{Results}

We propose and implement a novel numerical method to measure the 
configurational entropy, which fully resolves these issues.
The proposed methodology does not require precise definitions  
of a free energy landscape and metastable states. Our results show  
that the configurational entropy appears discontinuously 
at a characteristic low 
temperature, and that the Adam-Gibbs relation is not valid above 
the mode-coupling temperature. Therefore, this alternative approach 
provides a numerical estimate of the configurational entropy that is 
conceptually closer to theory, and yields quantitative results 
which agree better with experiments. 

The proposed 
numerical method is directly inspired by statistical mechanics approaches,
where the configurational entropy can be computed from the 
thermodynamic properties of constrained cloned 
systems~\cite{berthier_theoretical_2011,cavagna_supercooled_2009,monasson_structural_1995,mezard_thermodynamics_1999,mezard__????}. The physical idea is 
that constraining a system to reside `close' to a single state 
has a free energy cost $\Sigma(T)$, 
because it represents the entropic loss due to
an incomplete exploration of the configurational space.
To bypass the difficulty of defining metastable states rigorously, 
we obtain a numerical estimate of $\Sigma(T)$ by 
measuring a free energy difference between two 
thermodynamic phases that can be well-defined. 
In practice, we estimate $\Sigma(T)$  
from the thermodynamic properties of a system comprising 
two copies, 1 and 2, of the considered liquid thermalized at temperature $T$. 
As described in more detail in the Supplementary Information (SI), 
we conduct equilibrium simulations 
of these two coupled copies and carefully measure the probability distribution 
function of their mutual overlap,
$P(Q) = \langle \delta(Q - Q_{12}) \rangle$, where brackets indicate 
an equilibrium average. We define the overlap as 
$Q_{12} = N^{-1} \sum_{i,j}
\theta( a- | {\bf r}_{1,i} - {\bf r}_{2,j} | )$, 
where $\theta(x)$ is the Heaviside function, 
${\bf r}_{1,i}$ denotes the position 
of particle $i$ within configuration 1, $a$ is a length 
comparable to the particle diameter $\sigma$ (we take $a/\sigma = 0.3$), 
and $N$ the particle number in each copy.  
Note that this `collective' overlap is insensitive to particle exchanges. 
We define the 
`effective potential' $V(Q) = -\frac{T}{N} \ln P(Q)$, which is by definition 
the constrained equilibrium free energy of the total system 
when the average value of the overlap 
is $Q$~\cite{franz_phase_1997,cardenas_constrained_1999}. 

\begin{figure}[t]
\centerline{\includegraphics[width=.45\textwidth]{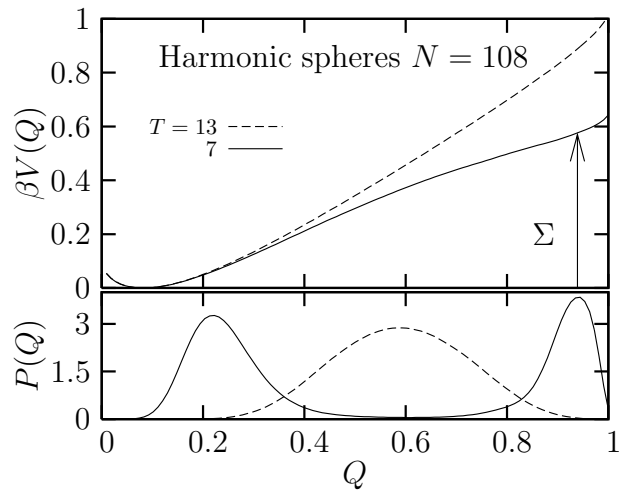}}
\caption{\label{fig1}Measurement of the configurational entropy defined 
in Eq.~(\ref{definition}), using the free energy difference
between the global minimum of $V(Q)$ at $Q_{\rm low}$ and the 
value at $Q_{\rm high}$, defined from the overlap distribution 
at coexistence. Top: Free energy $\beta V(Q)$ of $N=108$ harmonic spheres for 
two temperatures above (dashed line) and below (solid line) 
the critical temperature $T_c \approx 10$.
The arrow defines $\Sigma(T=7)$.
Bottom: The overlap distribution is bimodal below $T_c$ along the 
first-order transition line $\epsilon_c(T)$ (solid line), 
and featureless above $T_c$ (dashed line).}
\end{figure}

While $V(Q)$ was introduced long ago in theoretical 
calculations~\cite{franz_phase_1997}, 
it was only recently realized that it can be accurately 
determined in computer simulations by applying tools first devised 
to study equilibrium phase 
transitions~\cite{cammarota_phase-separation_2010,berthier_overlap_2013,parisi_liquid-glass_2013}. For a particular
model liquid, we have shown~\cite{berthier_overlap_2013} that 
$V(Q)$ is convex above 
a critical temperature $T_c$ below which it develops a linear part,
corresponding to a strongly non-Gaussian $P(Q)$. This observation 
implies that a thermodynamic field $\epsilon$ conjugated to the overlap
$Q$ induces, for $T < T_c$, an equilibrium first-order transition between 
a low-$Q$ and a high-$Q$ phase~\cite{franz_phase_1997}. This first-order 
transition line $\epsilon_c(T)$
ends at a second-order critical point at $T_c$, as explicitly demonstrated 
in~\cite{berthier_overlap_2013,parisi_liquid-glass_2013}. The existence 
of two phases below $T_c$ suggests to estimate 
the configurational entropy as:
\be
\Sigma(T) = \frac{1}{T} \left[  V(Q_{\rm high}) - V(Q_{\rm low})  \right], 
\label{definition}
\ee  
where $Q_{\rm low}$ denotes the position of the global minimum of $V(Q)$,
and $Q_{\rm high}$ is determined from the position of the peak in 
$P(Q)$ at coexistence, see Fig.~\ref{fig1}. 
Equation~(\ref{definition}) states that $\Sigma(T)$  
represents the free energy difference between the low-$Q$ phase where 
the two copies independently explore the free energy landscape 
and the high-$Q$ phase where they remain close to one another. 
This free energy difference originates from the fact that one of the 
copies cannot freely explore the configuration space,
and this precisely costs $\Sigma(T)$. (Details pertaining 
to quenched and annealed complexities are discussed below.)  
While the complexity in Eq.~(\ref{definition}) emerges naturally 
in mean-field calculations~\cite{franz_phase_1997}, 
our work is the first to implement 
this approach to estimate the configurational entropy in 
finite dimensional liquids.  

The definition (\ref{definition}) 
shows that the measurement of $\Sigma(T)$ does not rely 
on an explicit definition of a free energy landscape 
and of metastable states, and $\Sigma(T)$ does not
stem in the present approach from an enumeration of states. 
Instead, by measuring the thermodynamic properties of the high-$Q$
localized phase, we let the system itself define the extent of a `state'. 
This provides a direct determination of $\Sigma(T)$ which 
requires neither an approximate estimate of a vibrational 
contribution,  nor a detailed investigation of 
the potential energy landscape. This approach, which relies on the direct 
measurement of a free energy difference, is 
conceptually much closer to theoretical calculations.

Another consequence of Eq.~(\ref{definition}) 
is that $\Sigma(T)$ is only defined when two distinct phases 
can be distinguished, i.e. for $T \leq T_c$. 
For $T > T_c$, $V(Q)$ is featureless and $\Sigma(T)$ does not exist,
see Fig.~\ref{fig1}. In this regime, the entropy 
cannot be decomposed in configurational and vibrational parts.  
This is qualitatively consistent with specific theoretical 
calculations~\cite{Wolynes_1997,monasson_structural_1995,mezard_thermodynamics_1999,franz_phase_1997}. Physically, it means 
that the free energy landscape of the high temperature liquid 
has a simple topography for which the concept of configurational entropy 
is not relevant. A discontinuous emergence of the configurational 
entropy at low $T$ is naturally obtained within the 
present calculations, whereas it is
missed by previous methods~\cite{sciortino_inherent_1999,Sastry_2001,angelani_configurational_2007}. 

We studied two models of 
glass-formers using Monte-Carlo simulations~\cite{berthier_monte_2007}.
The first model is a 50:50 binary mixture of harmonic
spheres of diameter ratio 1.4~\cite{berthier_compressing_2009,berthier_glass_2009}.
Within reduced units~\cite{kob_non-monotonic_2012},
this quasi-hard sphere system has an onset temperature 
$T_{\rm on} \approx 12$~\cite{flenner_dynamic_2013}, a mode-coupling temperature $T_{\rm mct} 
\approx 5.2$~\cite{kob_non-monotonic_2012}, and a Vogel-Fulcher temperature 
$T_0 \approx 2$~\cite{flenner_dynamic_2013} (obtained with low reliability as
the system is weakly fragile at this density~\cite{berthier_compressing_2009,berthier_glass_2009}). 
We used $N=64$, 108 and 256, finding 
that finite size effects for $\Sigma(T)$ 
are small (see SI). We show data for $N=108$.
The second model is a 80:20 binary mixture of Lennard-Jones 
particles~\cite{kob_testing_1995}. In reduced 
units, the onset temperature is $T_{\rm on} \approx 1.0$, the mode-coupling
temperature $T_{\rm mct} \approx 0.435$~\cite{kob_testing_1995}, and the Vogel-Fulcher 
temperature $T_0 \approx 0.29$~\cite{Sastry_2001} (the model has 
intermediate fragility). We performed simulations 
with $N=180$. As described below (see Simulation Methods and SI
for detailed descriptions),
we combine umbrella sampling, multi-histogram reweighting and replica 
exachange techniques to quantity the rare fluctuations of the global 
overlap that need to be studied to obtain the free energy $V(Q)$. 
We find that differences between various 
possible estimates of $\Sigma(T)$~\cite{MPbook} can only be distinguished
in a very narrow temperature regime near $T_c$ which is not resolved 
by the present set of data, 
and therefore does not affect any of our conclusions.

\begin{figure}[t]
\centerline{\includegraphics[width=.45\textwidth]{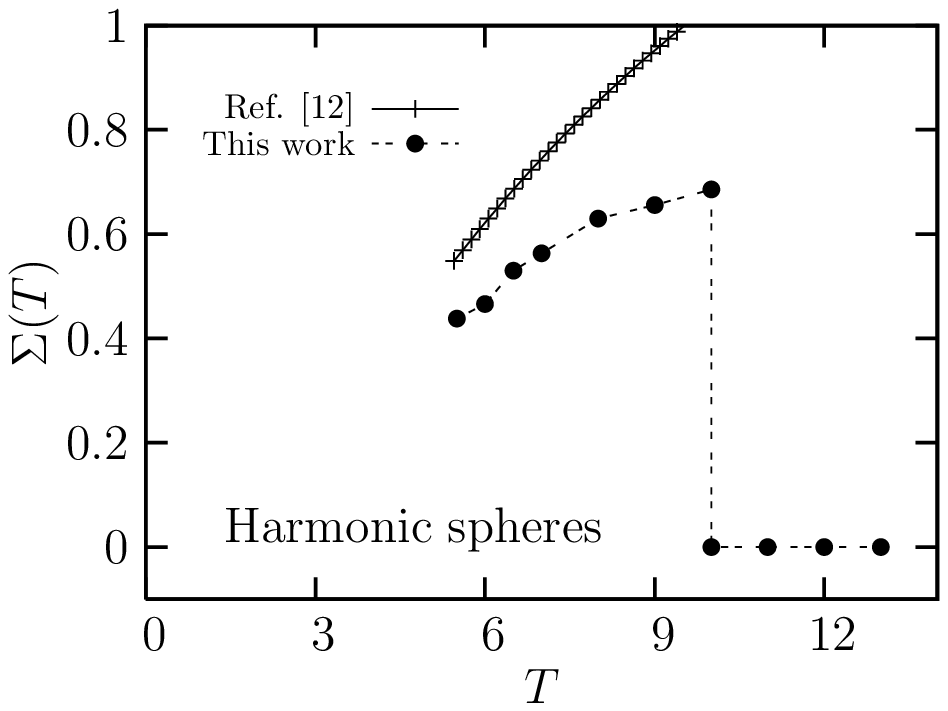}}
\centerline{\includegraphics[width=.45\textwidth]{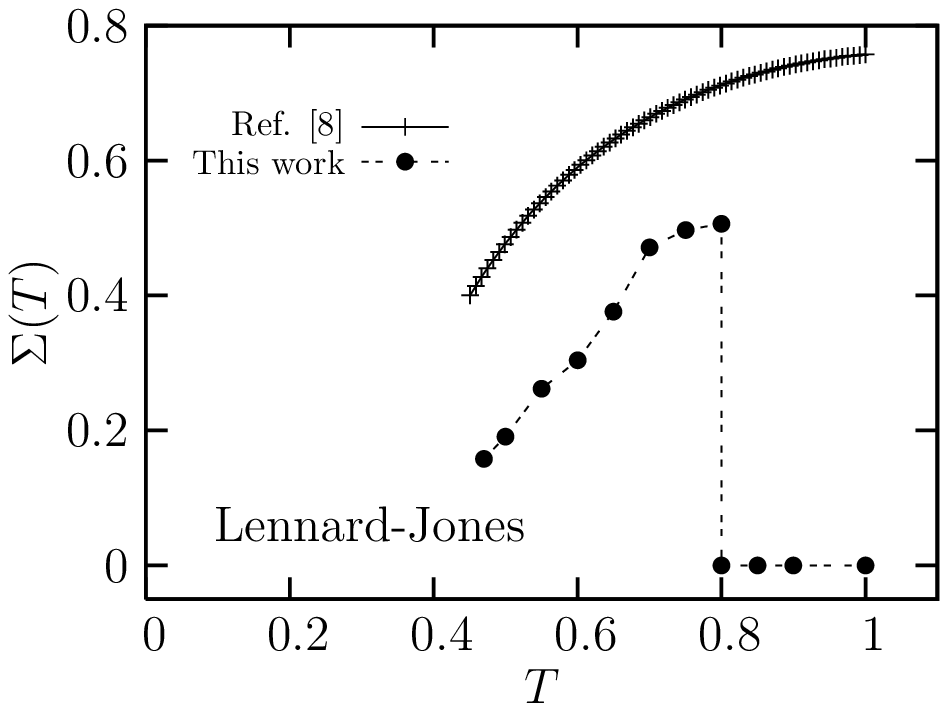}}
\caption{\label{fig2}The configurational entropy appears discontinuously
at temperatures $T_c \approx 10$ and $T_c \approx 0.8$, 
respectively for harmonic and Lennard-Jones particles,
in sharp contrast with literature data~\cite{angelani_configurational_2007,sciortino_inherent_1999}. 
(We used the mapping between hard and harmonic spheres 
discussed in \cite{berthier_compressing_2009,berthier_glass_2009} to convert the 
hard spheres data of \cite{angelani_configurational_2007} into equivalent data 
for harmonic spheres.) Note also the steeper 
temperature dependence of $\Sigma(T)$ 
in the more fragile Lennard-Jones model.}
\end{figure}

Our central results are in Fig.~\ref{fig2} which displays the
temperature dependence of $\Sigma(T)$ obtained from Eq.~(\ref{definition}) 
for two glass models.
In both cases, we find that $\Sigma(T)$ emerges discontinuously 
at a critical temperature $T_c$. We obtain $T_c \approx 10$ for 
harmonic spheres~\cite{berthier_overlap_2013}, 
and $T_c \approx 0.8$ for the Lennard-Jones model. 
Because $T_c$ is very close to, or slightly below, the onset 
temperature $T_{\rm on}$, this suggests 
that $T_c$ might represent a well-defined, physically meaningful 
definition of the onset temperature 
in supercooled liquids~\cite{sastry_signatures_1998}. Note that 
$T_c$ is significantly larger than $T_{\rm mct}$ 
obtained from a mode-coupling analysis of the dynamics.
While $T_c$ and $T_{\rm mct}$ are found to coincide in mean-field 
calculations~\cite{franz_phase_1997}, we find that $T_c$ remains well-defined 
in finite dimensions, whereas the mode-coupling singularity 
is replaced by a smooth crossover. 

The abrupt emergence of $\Sigma(T)$ at $T_c$ stands in sharp contrast 
with alternative methods~\cite{angelani_configurational_2007,sciortino_inherent_1999}, as demonstrated in Fig.~\ref{fig2}.
Therefore, the qualitative evolution of $\Sigma(T)$ obtained in 
this work is in closer agreement with theoretical and physical 
expectations (see e.g. Ref.~\cite{berthier_theoretical_2011}). 
Notice that such a discontinuous temperature dependence is of course
not observed experimentally, because experimental methods (just as previous
numerical methods) are not sensitive to the sharp emergence of metastable 
states that we are able to reveal here. Physically, 
our results simply suggest that a decomposition of the entropy in 
vibrational and configurational parts is not meaningful  
at high temperatures, a fact which is also hinted by the inherent 
structure approach~\cite{sastry_signatures_1998}.

The configurational entropy in Fig.~\ref{fig2} 
decreases steadily as temperature is lowered below $T_c$. 
This implies that the free energy difference between localized 
and delocalized states in configuration space decreases 
as $T$ gets lower, suggesting that the thermodynamic 
driving force to structural relaxation also decreases. 
A quantitative comparison with literature 
data in Fig.~\ref{fig2} shows that the temperature evolution of 
$\Sigma(T)$ below $T_c$ is in qualitative agreement with earlier work. 
However, the inherent structure formalism provides an estimate of the 
configurational entropy that is systematically larger than $\Sigma(T)$ 
over the explored range.
Despite the shortcomings mentioned above, inherent structure based 
approaches might still represent a valuable approximation at low $T$.

A motivation to determine $\Sigma(T)$ follows from Kauzmann's 
study of experimentally determined $\Sigma(T)$  
suggesting the existence of an entropy crisis, 
$\Sigma( T \to T_K) = 0$, possibly close to the 
Vogel-Fulcher temperature $T_0$~\cite{kauzmann_nature_1948}.
Our data do not cover a
broad enough temperature range to extrapolate an entropy crisis. However,
they do support the qualitative connection between thermodynamic and dynamic 
fragilities found experimentally~\cite{martinez_thermodynamic_2001}, 
since the more fragile
Lennard-Jones system also has a steeper $T/T_c$ 
dependence of $\Sigma(T)/\Sigma(T_c)$, as implied by Fig.~\ref{fig2}.
 
\begin{figure}[t]
  \begin{center}
\centerline{\includegraphics[width=.4\textwidth]{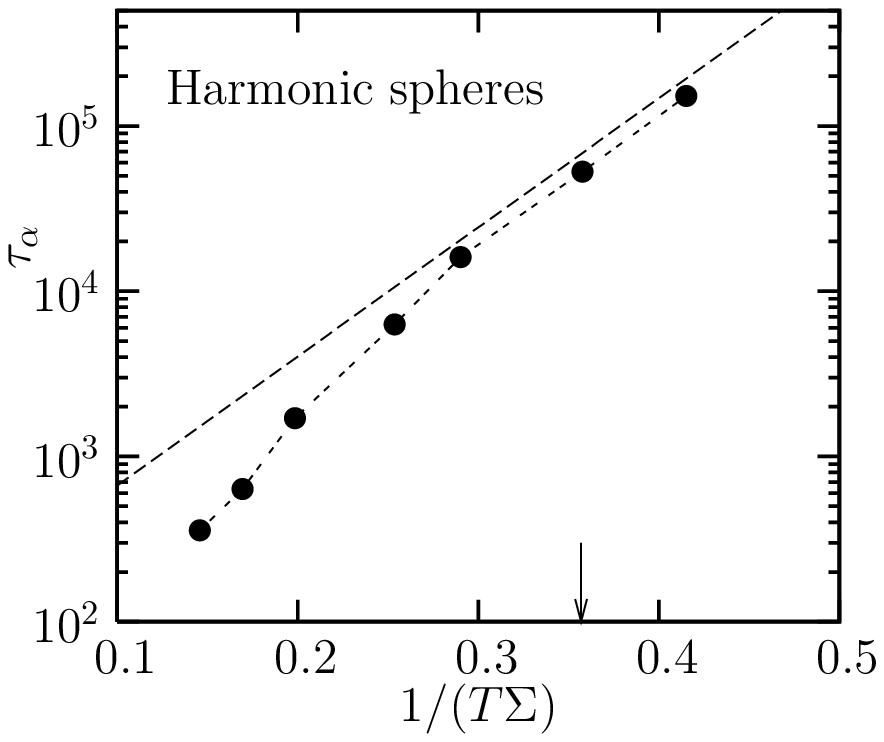}}
\centerline{\includegraphics[width=.38\textwidth]{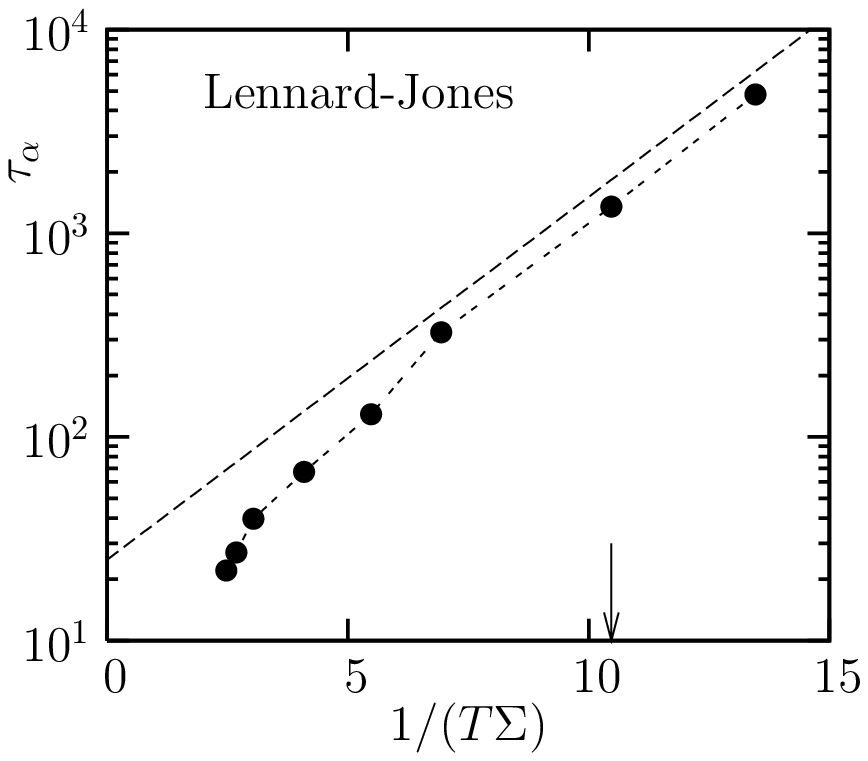}}  
  \end{center}
\caption{\label{fig3}Test of Adam-Gibbs relation in Eq.~(\ref{agr}).
The data show strong deviations from 
a linear relation between $\log \tau_\alpha$ and $(T \Sigma)^{-1}$ if 
temperature is not low enough,  whereas the relation 
is possibly satisfied when $T < 6$ (top) and $T < 0.5$ (bottom) (shown with
arrows), where the data seem to follow the indicated dashed lines.} 
\end{figure}

The Adam-Gibbs relation in Eq.~(\ref{agr}) is a quantitative
connection between thermodynamics and dynamics that can readily be 
tested once $\Sigma(T)$ is known, see Fig.~\ref{fig3}.
Notice first that, by construction, this relation cannot hold above
the critical temperature $T_c$ where $\Sigma(T)$ is not defined. 
Therefore, Eq.~(\ref{agr}) cannot be expected to work
if $T$ is too large. 
In fact, we find that it does not work well except close 
to $T_{\rm mct}$, although we would need more data 
to establish more firmly its validity at lower temperatures. Therefore, our 
results indicate that the broad range of validity of 
Eq.~(\ref{agr}) reported in earlier 
simulations~\cite{Scala_nature_2000,Sastry_2001,mossa_dynamics_2002,sengupta_dependence_2011} stems from using an alternative definition
of $\Sigma(T)$, for which Eq.~(\ref{agr}) holds over a broader 
range. We emphasize that our results conform to the 
general physical expectations that relaxation dynamics 
in supercooled liquids becomes thermally activated 
when temperature is low enough, typically below the mode-coupling 
temperature. The results exposed in Fig.~\ref{fig3} are 
therefore physically welcome, as there exists no 
fundamental reason for Eq.~(\ref{agr}) to be relevant
in the weakly supercooled temperature regime. 
Additionally, experiments find clear deviations from this relation 
in the temperature window covered by simulations~\cite{richert_dynamics_1998}. 
Therefore, our results suggest a plausible resolution to the existing
discrepancy between experiments and previous simulations, 
although more work remains to be done, especially at lower temperatures, 
to fully settle this issue.

\section{Discussion}

Despite the above successes, we emphasize that our 
determination of a configurational entropy $\Sigma(T)$ 
remains an approximation to the theoretical concept 
of a complexity counting the number of metastable states.
A first approximation stems from the fact that we
perform measurements of $V(Q)$ with two freely evolving 
copies. An alternative 
procedure~\cite{franz_phase_1997} 
consists of first drawing copy 1 from the equilibrium distribution, before 
studying the thermodynamics  of copy 2 in the presence of the quenched 
disorder imposed by copy 1.
This amounts to distinguishing annealed from quenched 
complexities~\cite{cardenas_constrained_1999}. In the mean-field 
limit where rigorous
calculations exist, the annealed $\Sigma(T)$ is 
an approximation to the quenched one, but the latter is more 
fundamental because it exactly counts the number of metastable
states. Although the quenched potential $V(Q)$ 
can also be measured~\cite{berthier_overlap_2013}, the procedure 
is more demanding. 
Before quantitatively comparing the two 
complexities, one should first establish more firmly 
the existence of a critical temperature $T_c$ in the quenched case. 
Preliminary work~\cite{berthier_overlap_2013} suggests a slight depression of 
the critical temperature $T_c$, and very close values for both 
$V(Q)$, but these issues need to be examined thoroughly.

A more fundamental issue concerns the
interpretation of $\Sigma(T)$ determined from $V(Q)$ 
as an entropy associated to the number of metastable states. This is
true at the mean-field level, where both $V(Q)$ and 
the complexity can be rigorously defined and 
computed~\cite{berthier_theoretical_2011,cavagna_supercooled_2009}.
The situation is ambiguous in finite dimensions, 
where infinitely long-lived metastable states do not exist, 
which forbids a strict definition of a complexity associated to their 
number~\cite{mezard__????}. 
Metastability can therefore only be approximately defined, 
for instance using finite timescales~\cite{biroli_metastable_2001} 
or lengthscales~\cite{bouchaud_adam-gibbs-kirkpatrick-thirumalai-wolynes_2004}, and metastable states cannot 
sharply emerge at the mode-coupling temperature, 
as they do in mean-field approximations~\cite{franz_phase_1997,mezard__????}. 
By contrast, we note that $V(Q)$ and $\Sigma(T)$ defined in 
Eq.~(\ref{definition}) do not suffer from these ambiguities. 
Therefore, our estimate of a configurational entropy is
well-defined in finite dimensions, even though metastable states are not.
This distinction also explains that a sharp emergence of $\Sigma(T)$ at $T_c$
is found in our simulations, whereas only a weak vestige of the
mode-coupling transition can be observed. 

We can tentatively interpret $\Sigma(T)$, as measured here,
as the entropy related to the number of `metastable' states, 
now defined over finite lengthscales,  
suggesting a possible deep connection between 
the emergence of $\Sigma(T)$ found here, 
and the growth of a static (point-to-set) correlation 
length~\cite{kob_non-monotonic_2012,bouchaud_adam-gibbs-kirkpatrick-thirumalai-wolynes_2004,biroli_thermodynamic_2008,berthier_static_2012,hocky_growing_2012}.
We emphasize that since $\Sigma(T)$ in Eq.~(\ref{definition}) 
quantifies the free energy cost to 
localize the system in configuration space, its rapid decrease 
upon supercooling is likely related to the slowing down of the dynamics.
This scenario naturally emerges in thermodynamic theories of the 
glass transition, such as Adam-Gibbs and random 
first-order transition theories.
The numerical strategy proposed herein provides a sensible measure of 
the configurational entropy in 
the temperature range currently accessible to numerical simulations and 
will thus allow a stringent test of
theoretical approaches in which the configurational entropy plays a 
central role.

\section{Materials and methods}

{\bf Models --} 
Our first model is a binary mixture of harmonic 
spheres~\cite{berthier_compressing_2009,berthier_glass_2009},
where particles of type  $\alpha,\beta$ interact by a 
harmonic potential, 
$$V_{\alpha\beta}(r) = \epsilon_{\alpha \beta}
(1  -  r / \sigma_{\alpha \beta})^2,$$  which is truncated 
at distance $r=\sigma_{\alpha \beta}$.  
The particle types are labeled $A$ and $B$ and the interaction
parameters are $(\sigma_{AA},\sigma_{AB},\sigma_{BB})=
(1.0,1.2,1.4) \, \sigma$, and $(\epsilon_{AA},\epsilon_{AB},
\epsilon_{BB})=(1.0,1.0,1.0) \, \varepsilon$.  
We consider $N=64$, 108 and 256 with $N_{A} = N_{B} = N/2$.
We express length scales in units of $\sigma$, temperatures
in units of $10^{-4} \varepsilon$, and perform simulations 
at constant number density $\rho = 0.675$.  

In the Kob-Andersen mixture, particles of types $\alpha,\beta$ 
interact by a Lennard-Jones potential $$V_{\alpha\beta}(r) = 
4\epsilon_{\alpha\beta}[(\sigma_{\alpha\beta}/r)^{12} - 
(\sigma_{\alpha\beta}/r)^6],$$ which is
truncated and shifted at $r=2.5\sigma_{\alpha\beta}$.
The particle types are labelled $A$ and $B$ and the interaction
parameters are $(\sigma_{AA},\sigma_{AB},\sigma_{BB})=
(1.0,0.80,0.88)\, \sigma$
and $(\epsilon_{AA},\epsilon_{AB},\epsilon_{BB})=(1.0,1.5,0.5)
\, \varepsilon$.  We consider $N=180$ particles with 
$N_{A} = 4N/5$ and $N_{B} = N/5$. We express length scales in 
units of $\sigma$, temperatures in units of $\varepsilon$, 
and perform simulations at constant number density $\rho = 1.2$.  

{\bf Simulation methods--}
Both models are studied numerically using Monte-Carlo
simulations~\cite{berthier_monte_2007}. 
Measuring $\Sigma(T)$ from Eq.~(\ref{definition}) is numerically 
challenging as it requires the determination of $V(Q)$ over a broad 
range of $Q$, which necessitates a quantitative analysis of atypical
overlap fluctuations. This difficulty 
is efficiently overcome by using umbrella sampling 
techniques~\cite{frenkel_understanding_2001}. Briefly, $V(Q)$ is obtained 
by gathering the results of a series of $n$ simulations
biased in such a way that distinct simulations explore 
distinct ranges of overlap values~\cite{berthier_overlap_2013}. 
Histogram reweighting techniques are then used to reconstruct 
$P(Q)$ over the complete $Q$ range~\cite{berthier_overlap_2013}. Another 
challenge is the difficulty to ensure proper thermalization of 
each simulation, 
which becomes serious when $Q$ is large and $T$ is low. 
This is not prohibitive in studies where only the 
vicinity of the critical temperature $T_c$ is 
explored~\cite{berthier_overlap_2013,parisi_liquid-glass_2013}.
To access much lower temperatures,  
we have introduced replica-exchange 
Monte-Carlo moves between the $n$ biased simulations, 
borrowing techniques used in phase transition 
studies~\cite{frenkel_understanding_2001,hukushima_exchange_1996,yan_hyper-parallel_1999}. 
Because each simulation now performs a random walk in 
parameter space, thermalization is greatly enhanced, and 
lower temperatures can be sampled. 
By approaching the mode-coupling temperature, we are able to measure 
$\Sigma(T)$ over a physically significant $T$-range. 
The procedure can presumably be further optimized to access even 
lower temperatures. It can also easily be 
applied to different models, including hard spheres to which 
the inherent structure formalism does not 
apply~\cite{angelani_configurational_2007}.
Full details about these methods as well as about finite size 
effects are given in the SI.

\begin{acknowledgments}
We thank G. Biroli, G. Parisi and G. Tarjus for useful exchanges.
The research leading to these results has received funding
from the European Research Council under the European Union's Seventh
Framework Programme (FP7/2007-2013) / ERC Grant agreement No 306845.
\end{acknowledgments}

\newpage

\begin{appendix}

\section{Supplementary information}

In this appendix, %we provide details about the 
%two numerical models of glass-forming liquids studied in this work, 
we describe the umbrella sampling, parallel tempering 
and histogram reweighting techniques we used to measure 
numerically the configurational entropy defined by Eq.~(2),
and discuss finite size effects.  

\subsection{Umbrella sampling}

To access the probability distribution function of the overlap 
at a given temperature, we conduct $n$ distinct simulations.
In each simulation, $i = 1, \cdots, n$, two copies of the system
evolve according to the following Hamiltonian:
\begin{equation}
H_i = H [ \{ {\bf r_1} \}]  + H [ \{ {\bf r_2} \} ] 
- \epsilon Q_{12} +  W_i(Q_{12}), 
\label{hi}
\end{equation}
where $H [\{ {\bf r} \} ]$ is the Hamiltonian of the original liquid, 
$\{ {\bf r_1} \}$ and $\{ {\bf r_2} \}$ respectively 
represent the positions of the $N$ particles in copies 1 and 2, 
$Q_{12}$ is the overlap between configurations 1 and 2,  
$\epsilon$ is the thermodynamic field conjugated to the overlap, 
and the biasing potential $W_i(Q)$ is taken of the form:
\begin{equation}
W_i(Q) = k_i (Q - Q_i)^2,
\label{wi}
\end{equation} 
with parameters $(k_i, Q_i)$ chosen to constrain the 
overlap $Q$ to explore values away from its average equilibrium value. 

We perform simple Monte-Carlo moves, where we attempt 
single particle displacements of small amplitude~\cite{SIMC}
(typically of size 0.05 $\sigma$)
which we accept using a Metropolis acceptance rate given by 
Hamiltonian (\ref{hi}). 
We define time steps such that one Monte-Carlo step represents 
$2 \times N$ attempts to displace randomly chosen particles among the two copies
of the system.

Provided it is properly thermalized (see below),
the main outcome of a given simulation is the probability distribution 
function of the overlap, 
\begin{equation}
P_i(Q,\epsilon,T) = \langle \delta(Q - Q_{12}) \rangle_i,
\end{equation} 
where $\langle \cdots \rangle_i$ represents a thermal average 
with the Hamiltonian $H_i$ in Eq.~(\ref{hi}) for a
given state point defined by $(\epsilon, T)$. 

The idea behind the biasing potentials $W_i(Q)$ in Eqs.~(\ref{hi},
\ref{wi})
is that the fluctuations of the overlap in each simulation can be
tailored to explore a narrow region centered around $Q_i$. 
Therefore, each simulation explores only a small range of 
overlap values, and it becomes unnecessary to wait for 
very rare overlap fluctuations to occur. 
Therefore, umbrella sampling is the key method to 
efficiently measure atypical overlap fluctuations~\cite{SIumbrella}. 

\begin{figure}
\psfig{file=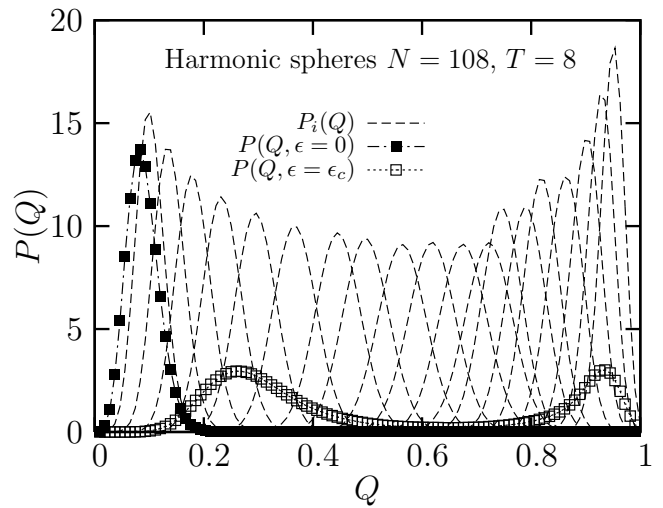,width=8.5cm}
\caption{Numerical measurement of $V(Q)$ for harmonic spheres
with $N=108$ and $T=8$. We perform $n=19$ simulations in parallel with distinct
biasing potentials $W_i(Q)$ chosen to explore the entire range of overlap
between 0 and 1. Each simulation returns a distribution $P_i(Q)$ (dashed 
lines). A significant overlap between neighboring distributions
ensures the efficiency of the replica-exchange moves.  
Using multi-histogram reweighting technique, the biased 
distributions can be gathered to provide the equilibrium (unbiased)
overlap distribution either at $\epsilon = 0$ (closed symbols), 
from which $V(Q)$ is directly obtained, or at finite field $\epsilon$. 
At the critical field $\epsilon_c(T)$, the equilibrium distribution is bimodal 
(open symbols), reflecting phase coexistence between localized and 
delocalized states.} 
\label{figSI1}
\end{figure}

In Fig.~\ref{figSI1} we show the distributions $P_i(Q,\epsilon,T)$
measured in $n=19$ simulations of 
harmonic spheres at $T=8$ for $N=108$ for a 
given set of biasing potentials chosen 
to adequately cover the range of overlap between 0 and 1. We  
used $Q_i$ values in the interval $[-0.08, 0.93]$, and 
a strength of the Gaussian trap $k_i$ in the interval $[2.8, 4.6]$.  
We observe that each simulation returns a probability distribution function 
which is relatively narrow and approximately Gaussian, showing
that the sampling of each overlap sector is no more slowed down 
by the emergence of non-trivial distributions of overlap 
fluctuations~\cite{SIumbrella}. In other words, simulations are faster 
because the Hamiltonian $(\ref{hi})$ has no phase transition 
in the $(\epsilon, T)$ plane. 

\subsection{Replica-exchange}

While the umbrella sampling technique described above considerably accelerates 
the measurement of the overlap fluctuations, we have observed 
that when $T$ is low, 
$N$ is large, and/or $Q_i$ is large, the particle 
dynamics slows down considerably, and it becomes
difficult to perform an accurate sampling of the overlap 
fluctuations imposed the Hamiltonian (\ref{hi}), because the overlap
fluctuations become slow.  
This sampling problem was already mentioned in 
Ref.~\cite{SIpre}, and this prevented the exploration of
temperatures much lower than $T_c$. 

To accurately explore the temperature regime needed for the present work, 
we have implemented replica-exchanges Monte-Carlo moves~\cite{SIPT}. 
We now conduct 
the $n$ simulations needed for the umbrella sampling 
at temperature $T$ in parallel, and propose Monte-Carlo exchange moves 
between neighboring simulations characterized by nearby sets of parameters,
say $(k_i, Q_i)$ and $(k_{i+1}, Q_{i+1})$. An exchange between 
simulations $i$ and $i+1$ is proposed 
with a low frequency (typically every $10^4$ Monte-Carlo steps)
and they are accepted with a Metropolis acceptance rate 
given by the Hamiltonians $H_i$ and $H_{i+1}$, ensuring 
that simulations satisfy detailed balance. 

Because each simulation now performs a random walk in the 
parameter space defined by $\{ (k_i, Q_i), \, i=1, \cdots, n \}$, 
the sampling of the overlap 
fluctuations is greatly enhanced, even for the `hard' cases.
For the method to be efficient, we need to adjust the 
biasing potentials $W_i(Q)$ such that the overlap distributions
in each simulation largely overlap, as can be seen in 
the example shown in Fig.~\ref{figSI1}. We have used up to 
$n=24$ parallel simulations to gather our data. We have carefully checked 
thermalization by running simulations for very long times
(up to $5 \times 10^9$ Monte-Carlo steps per simulation), 
making sure that each state point was visited several times 
by all simulations due to the replica-exchange. This 
represents a significant numerical effort. 
 
\subsection{Histogram reweighting}

Having obtained thermalized results from $n$ biased simulations
running in parallel, we process the simulation outcome 
using multi-histogram reweighting 
methods to reconstruct the unbiased probability $P(Q)$ 
from the $n$ independently measured $P_i(Q)$~\cite{SIumbrella,SIpre}, 
\be
P(Q,\epsilon,T) 
= \frac{\sum_{i=1}^n P_i(Q,\epsilon,T) }{ \sum_{i=1}^n e^{-\beta W_i} / Z_i },
\label{pq}
\ee 
where the $Z_i$ are defined self-consistently as 
\be
Z_i = \int_0^1 dQ' \frac{ \sum_{j=1}^n P_j(Q',\epsilon,T) }{ \sum_{j=1}^n 
e^{\beta(W_i-W_j) }/Z_j }.
\ee 

Notice that the value of $\epsilon$ used in the simulations plays no 
conceptual role because the reweighting method allows us to 
directly obtain $P(Q,\epsilon',T)$ from $P(Q,\epsilon,T)$
for distinct field values $\epsilon$ and $\epsilon'$:
\begin{equation}
P(Q,\epsilon',T ) = \frac{P(Q,\epsilon,T) e^{-\beta Q 
(\epsilon'-\epsilon)}}{\int_0^1 dQ' P(Q',\epsilon,T) e^{-\beta Q' 
(\epsilon'-\epsilon)} } . 
\end{equation}
We have applied the combined umbrella sampling / replica-exchange 
technique~\cite{SIyan}
both with $\epsilon=0$ for the Lennard-Jones potential, 
and with $\epsilon$ adjusted to be roughly at phase coexistence 
for the harmonic sphere system. We found that the latter method 
allows for an easier convergence of the simulation parameters, without
affecting qualitatively the efficiency of statistical sampling.   

Two values of the field $\epsilon$ are particularly relevant for this 
study. First, we obtain the potential $V(Q)$ as: 
\begin{equation}
V(Q) = - \frac{T}{N} \ln P(Q, \epsilon=0, T). 
\end{equation}
Notice that $V(Q)$ is only defined up to an additive constant, 
which we adjust such that $V (Q_{\rm low}) = 0$, where
$Q_{\rm low}$ is defined as the location of the global minimum in 
$V(Q)$. This additive constant is irrelevant, as we only need 
to determine free energy differences to determine $\Sigma(T)$.
In Fig.~\ref{figSI1} we show the distribution 
$P(Q, \epsilon=0, T)$ for the chosen example. The potential $V(Q)$
in Fig.~1 is simply obtained by taking (minus) 
the logarithm of this function which allows for a clearer view of the tail
of the distribution. 

A second useful quantity is the overlap distribution obtained 
in the presence of a field $\epsilon$ ensuring phase coexistence 
below $T_c$. In practice, we use the strength of the reweighting 
method to finely explore a range of $\epsilon$ values, 
and define $\epsilon_c(T)$ as the field for which 
the distribution $P_i(Q,\epsilon_c,T)$ yields a maximum amount of 
fluctuations (as quantified by their variance). This corresponds 
to the situation where the two peaks of the distribution 
have nearly equal height. 
From this bimodal distribution we define the 
position of the high-overlap value $Q_{\rm high}$, as illustrated in 
Fig.~1. 

It is useful to notice that $\epsilon_c$ 
has a simple graphical interpretation, since it represents 
by definition the amplitude of the field needed to `tilt' the 
potential $V(Q)$ towards coexistence, see Fig.~1.
Therefore, it represents a simple proxy to the 
free energy difference $\Sigma(T)$ defined in Fig.~1, 
because the relation
\begin{equation}
\Sigma \approx Q_{\rm high} \, \epsilon_c,
\label{approx}
\end{equation}
holds to a good approximation. We find that differences between various 
possible estimates of $\Sigma(T)$~\cite{SIMPbook} can only be distinguished
in a very narrow temperature regime near $T_c$ which is not resolved 
by the present set of data, 
and therefore does not affect any of our conclusions.
This estimate is also useful because it stems from a 
single number (the critical field $\epsilon_c$), 
and is less prone to statistical errors than the 
entire function $V(Q)$. 

We also remark that, contrary to mean-field calculations, 
$Q_{\rm high}$ cannot be defined from the existence
of a secondary minimum in $V(Q)$, because the potential 
has to be a convex function of $Q$ in finite dimensions.
Therefore, a secondary minimum cannot exist in the 
large system size limit in our simulations. This is why
we have instead defined $Q_{\rm high}$ from the position 
of the large-$Q$ peak in $P(Q)$ at coexistence. 

In the same vein, the appearance in mean-field 
calculations of a secondary minimum 
in $V(Q)$ is the direct signature of the mode-coupling 
transition at $T=T_{\rm mct}$. Because no local minimum 
appears in $V(Q)$ at any temperature in finite dimensions,
we do not expect to detect any qualitative change in 
$V(Q)$ in the region of the mode-coupling temperature 
in our simulations. This immediately suggests that $T_{\rm mct}$ can not 
be expected to play any significant role in the
present study. 

\subsection{Finite size effects}

\begin{figure}
\psfig{file=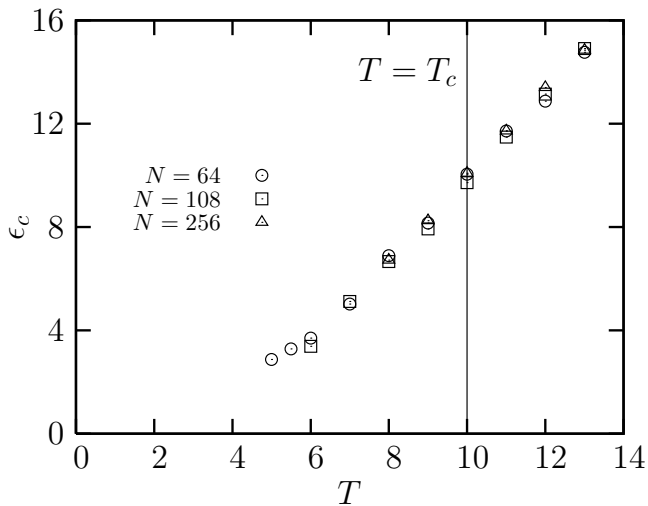,width=8.5cm}
\caption{Study of finite size effects in harmonic spheres, using 
the value of the critical field $\epsilon_c$ as an approximation
to the configurational entropy below $T_c$, showing no clear trend 
with $N$, within statistical accuracy.}
\label{figSI2}
\end{figure}

For the harmonic sphere model we have used 3 different 
system sizes. This allows us to discuss how finite size effects 
affect our determination of the configurational entropy
in this system.

Because our definition of a configurational 
entropy results from the existence of an underlying phase transition
for the constrained Hamiltonian (\ref{hi}), finite size effects
should naturally be carefully discussed.
However, while a strong system size dependence might be expected 
to affect the overlap fluctuations near the critical temperature $T_c$, 
much smaller finite size effects can be expected in the low-temperature 
regime $T< T_c$ of interest here. Therefore a finite size system 
could potentially affect the value of $Q_{\rm high}$ and the 
free energy difference between the two phases, but these can be 
expected to be small, just as finite size effects would also be 
small in the magnetic phase of the Ising model. 

To test this idea, we present in Fig.~\ref{figSI2} the 
value of the critical field $\epsilon_c$ for different system
sizes and temperatures in the harmonic sphere model, which 
represents a faithful estimate to the configurational entropy,
see Eq.~(\ref{approx}). 
The data shown in Fig.~\ref{figSI2} clearly indicate that finite size 
effects are small in the temperature regime explored by the
present simulations. We find similarly that the value of $Q_{\rm high}$
is not strongly affected by finite size effects.
We expect that a stronger influence 
of the system size could be 
observed at much lower temperatures, when the system size 
competes more strongly with the growing point-to-set correlation 
length of the system, which is found to be relatively modest 
in the temperature regime above the mode-coupling 
transition~\cite{SIpts3}.

\end{appendix}

\end{document}